\documentclass[a4paper]{article}
\usepackage{listings}
\usepackage{xcolor}
\usepackage{multirow}
\IfFileExists{jheppub.sty}{\usepackage{jheppub}}{}
\usepackage{dcolumn}
\usepackage[english]{babel}
\usepackage[utf8x]{inputenc}
\usepackage[T1]{fontenc}
\usepackage{palatino}
\usepackage{amsmath}
\usepackage{afterpage}
\usepackage{graphicx, subcaption}
\usepackage[colorinlistoftodos]{todonotes}
\usepackage[colorlinks=true]{hyperref}
\usepackage{makeidx}
\usepackage[switch]{lineno}
\newcommand{\be}{\begin{equation}}  
\newcommand{\ee}{\end{equation}}  
\newcommand{\bea}{\begin{eqnarray}}  
\newcommand{\eea}{\end{eqnarray}}  
\makeindex
\begin{document}

\vspace*{1.2cm}

\thispagestyle{empty}
\begin{center}

{\LARGE \bf Study of the thermodynamic properties of hot QCD matter with the CMS experiment}

\par\vspace*{7mm}\par

{

\bigskip

\large \bf Cesar A. Bernardes on behalf of the CMS Collaboration~\footnote{Copyright 2026 CERN for the benefit of the CMS Collaboration. Reproduction of this article or parts of it is allowed as specified in the CC-BY-4.0 license.}}

\bigskip

{\large \bf cesar.bernardes@cern.ch}

\bigskip

{Núcleo de Computação Científica (NCC) - UNESP}

\bigskip

{\it Presented at the Workshop of Advances in QCD at the LHC and the EIC, CBPF, Rio de Janeiro, Brazil, November 9-15 2025}

\vspace*{15mm}

\end{center}
\vspace*{1mm}

\begin{abstract}

These proceedings summarize recent CMS measurements at the LHC that extract the squared speed of sound, \(c_s^2\), of strongly interacting matter at extreme temperatures from the multiplicity dependence of the mean transverse momentum in ultra-central lead-lead (PbPb) collisions at \(\sqrt{s_{\mathrm{NN}}} = 5.02\ \mathrm{TeV}\). The analysis yields \(c_s^2 = 0.241 \pm 0.002\, (\mathrm{stat}) \pm 0.016\, (\mathrm{syst})\) at an effective temperature of \(T_{\mathrm{eff}} = 219 \pm 8\, (\mathrm{syst})\ \mathrm{MeV}\), in good agreement with lattice-QCD calculations. Complementary studies in proton-lead (pPb) collisions are also presented to investigate possible quark-gluon plasma signatures in smaller systems. 

\end{abstract}

 \section{Introduction}

Ultrarelativistic heavy-ion collisions create an extreme state of matter known as the quark-gluon plasma (QGP), in which quarks and gluons are deconfined over extended volumes~\cite{Shuryak:1978ij}. This strongly interacting medium exhibits remarkable collective behavior with minimal viscous dissipation, behaving as an almost ``perfect liquid''~\cite{STAR:2004gwh,PHENIX:2004wyw,BRAHMS:2004vtz,PHOBOS:2004eai}. The macroscopic properties of the QGP are usually well described by relativistic hydrodynamics, where the equation of state (EoS), e.g., the relationship between pressure \(P\) and energy density \(\epsilon\), plays a fundamental role. 

A key quantity governing the EoS is the speed of sound, \(c_s\), defined as

\begin{equation}
c_s^2 = \left.\frac{\mathrm{d}P}{\mathrm{d}\epsilon}\right|_{\mathrm{adiabatic}} = \frac{s\,\mathrm{d}T}{T\,\mathrm{d}s},
\label{eq:1}
\end{equation}

where \(s\) is the entropy density and \(T\) is the temperature~\cite{LandauLifshitz} of the system in equilibrium. This parameter characterizes how the medium responds to compression and expansion, and its measurement provides stringent constraints on the underlying degrees of freedom of QCD matter.

This proceedings presents a measurement of \(c_s^2\) by the CMS experiment using ultra-central PbPb collisions~\cite{CMS:2024kxg}, following a novel methodology proposed in Ref.~\cite{Gardim:2019nnr}. The approach exploits the fact that in collisions with near-zero impact parameter (ultra-central collisions), the system volume is approximately fixed by the nuclear size, while event-by-event fluctuations in the number of interacting partons allow the energy density to vary. Under these conditions, the thermodynamic relation

\begin{equation}
c_s^2 = \frac{\mathrm{d}\ln\langle p_{\mathrm{T}}\rangle}{\mathrm{d}\ln N_{\mathrm{ch}}}
\label{eq:2}
\end{equation}

can be applied, where \(\langle p_{\mathrm{T}}\rangle\) is the mean transverse momentum of charged particles (proportional to the effective temperature) and \(N_{\mathrm{ch}}\) is the charged-particle multiplicity (proportional to the entropy)~\cite{Gardim:2019brr,Gardim:2019nnr}. In this context, \(c_s^2\) should not be interpreted as the speed of sound of the QGP produced in heavy-ion collisions; rather, it corresponds to the speed of sound of an effective system in global thermal equilibrium, with total energy and entropy matched to their average values on a hypersurface around freeze-out, including the kinetic contribution from medium expansion~\cite{Gardim:2019brr}. 

We also present first measurements in pPb collisions~\cite{CMS:HIN25001}, exploring whether similar thermodynamic properties can be extracted in smaller systems where QGP formation is still under active investigation. 
 
\section{Experimental method}

\subsection{The CMS detector}

The CMS apparatus~\cite{CMS:2008xjf} is a multipurpose, nearly hermetic detector designed to trigger on and identify electrons, muons, photons, and hadrons. Key detector components for this analysis include the silicon pixel and strip tracker (providing precise charged-particle tracking within \(|\eta| < 2.4\)), the hadron forward (HF) calorimeters (extending coverage to \(3.0 < |\eta| < 5.2\)), and the zero-degree calorimeters (ZDCs) used for pileup rejection in ultra-central PbPb collisions.

\subsection{Data samples and event selection}

The PbPb analysis uses data collected in 2018 at \(\sqrt{s_{\mathrm{NN}}} = 5.02\ \mathrm{TeV}\), corresponding to an integrated luminosity of \(0.607\ \mathrm{nb}^{-1}\). Minimum-bias events are triggered by requiring energy deposits above threshold in both HF calorimeters. Pileup events (concurrent interactions per bunch crossing) are rejected using correlations between HF and ZDC energy deposits. The centrality, degree of overlap between the two colliding nuclei, is determined from the total transverse energy in both HF calorimeters, \(E_{\mathrm{T,sum}}^{\mathrm{HF}}\)~\cite{CMS:2024kxg}. 

For pPb collisions, data at \(\sqrt{s_{\mathrm{NN}}} = 5.02\ \mathrm{TeV}\) (\(0.51\ \mathrm{nb}^{-1}\)) and \(8.16\ \mathrm{TeV}\) (\(186\ \mathrm{nb}^{-1}\)) recorded in 2016 are used. High-multiplicity triggers (based on the calorimeters and the tracker) enable collection of events with large final-state particle yields, where signatures of collectivity is observed. The pileup events are rejected by applying a selection based on the distance between primary and secondary vertices and their associated charged particle multiplicities~\cite{CMS:HIN25001}. 

\subsection{Track reconstruction and corrections}

Track reconstruction for central PbPb events is performed using two algorithms to manage computational load. The first method reconstructs tracks with \(p_{\mathrm{T}} > 1.0\ \mathrm{GeV}\) using both pixel and strip tracker hits; the second reconstructs tracks with \(0.3 < p_{\mathrm{T}} < 1.0\ \mathrm{GeV}\) using only the pixel detector. Tracks are selected with \(|\eta| < 0.5\) for optimal performance. Tracking efficiency \(\epsilon_{\mathrm{eff}}\) and misreconstruction rate \(\epsilon_{\mathrm{mis}}\) are evaluated using HYDJET~\cite{Lokhtin:2008xi} events with full GEANT4~\cite{Agostinelli:2002hh} detector simulation. A correction factor \(\epsilon_{\mathrm{trk}} = (1-\epsilon_{\mathrm{mis}})/\epsilon_{\mathrm{eff}}\) is applied as a function of \(\eta\), \(p_{\mathrm{T}}\), and detector occupancy (estimated from pixel cluster counts).

For pPb collisions, just one method using both pixel and strip tracker hits is used. Selected tracks with \(|\eta| < 1.5\) and \(p_{\mathrm{T}} > 0.3\ \mathrm{GeV}\) are used~\cite{CMS:HIN25001}, and a similar correction as applied to PbPb is used.

\subsection{Observable definition}

The primary observable is the mean transverse momentum \(\langle p_{\mathrm{T}}\rangle\) as a function of charged-particle multiplicity \(N_{\mathrm{ch}}\), both measured within the same kinematic acceptance. Charged-particle \(p_{\mathrm{T}}\) spectra for \(p_{\mathrm{T}} > 0.3\ \mathrm{GeV}\) are measured in bins of \(E_{\mathrm{T,sum}}^{\mathrm{HF}}\). To avoid bias, spectra are extrapolated to the full \(p_{\mathrm{T}}\) range by fitting a Hagedorn function~\cite{Hagedorn:1983wk} to the measured range \(0.4 < p_{\mathrm{T}} < 4.5\ \mathrm{GeV}\) (PbPb) or \(0.3 < p_{\mathrm{T}} < 1.5\ \mathrm{GeV}\) (pPb). Normalized quantities \(\langle p_{\mathrm{T}}\rangle^{\mathrm{norm}} = \langle p_{\mathrm{T}}\rangle/\langle p_{\mathrm{T}}\rangle^{0}\) and \(N_{\mathrm{ch}}^{\mathrm{norm}} = N_{\mathrm{ch}}/N_{\mathrm{ch}}^{0}\) are used to minimize systematic uncertainties, where \(\langle p_{\mathrm{T}}\rangle^{0}\) and \(N_{\mathrm{ch}}^{0}\) are values in a reference event class (0--5\% centrality for PbPb, 0--0.01\% for pPb).

The speed of sound is extracted by fitting the relation~\cite{Gardim:2019nnr}

\[
\langle p_{\mathrm{T}}\rangle^{\mathrm{norm}} = \left(\frac{N_{\mathrm{ch}}^{\mathrm{norm}}}{\langle N_{\mathrm{ch}}^{\mathrm{knee}}|N_{\mathrm{ch}}^{\mathrm{norm}}\rangle}\right)^{c_s^2},
\]

where \(\langle N_{\mathrm{ch}}^{\mathrm{knee}}|N_{\mathrm{ch}}^{\mathrm{norm}}\rangle\) accounts for the knee-shaped multiplicity distribution at \(b=0\)~\cite{Das:2017ned}. For pPb collisions, a two-energy method is employed~\cite{Gardim:2019brr}, fitting \(\langle p_{\mathrm{T}}\rangle = C N_{\mathrm{ch}}^{c_s^2}\) at fixed centrality intervals across \(\sqrt{s_{\mathrm{NN}}} = 5.02\) and \(8.16\ \mathrm{TeV}\). Recent theoretical studies have identified limitations in these methodologies~\cite{Gavassino:2025myk, Mu:2025kgl} and have also motivated new measurements.

Dominant systematic uncertainties arise from tracking corrections and \(p_{\mathrm{T}}\) extrapolation. Tracking uncertainties are evaluated by varying track selections; \(p_{\mathrm{T}}\) extrapolation uncertainties by varying fit ranges and using alternative fit functions. For normalized quantities, most uncertainties cancel significantly. Total systematic uncertainties are obtained by adding individual contributions in quadrature.

\section{Results}

\subsection{Speed of sound in PbPb collisions}

\begin{figure}[!ht]
\begin{center}
\includegraphics[width=0.80\textwidth]{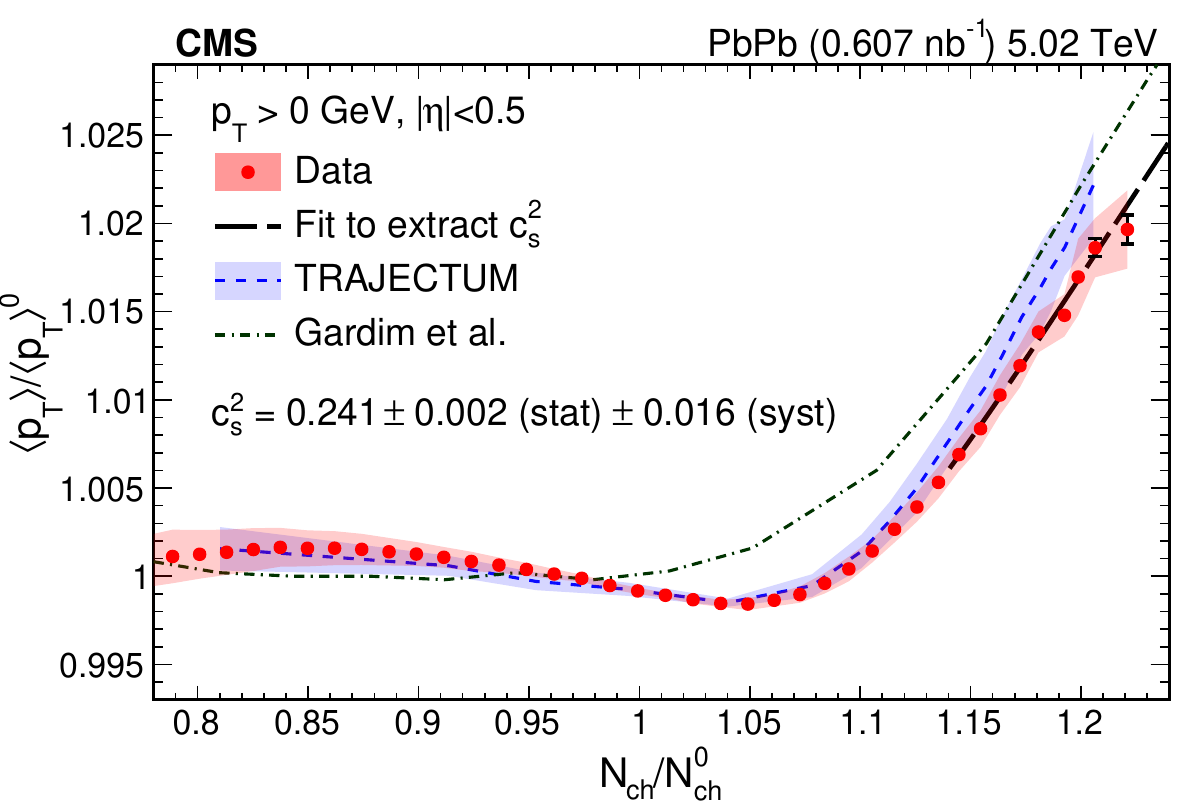}
\caption{The normalized mean transverse momentum of charged particles as a function of normalized charged-particle multiplicity, measured within $|\eta|<0.5$ and extrapolated to the full $p_T$ range. Hydrodynamic-simulation results are also shown for comparison~\cite{CMS:2024kxg}.}
\label{fig:1}
\end{center}
\end{figure}

Figure~\ref{fig:1} shows \(\langle p_{\mathrm{T}}\rangle^{\mathrm{norm}}\) as a function of \(N_{\mathrm{ch}}^{\mathrm{norm}}\) for PbPb collisions at \(\sqrt{s_{\mathrm{NN}}} = 5.02\ \mathrm{TeV}\). A weak decline toward a local minimum near \(N_{\mathrm{ch}}^{\mathrm{norm}} \sim 1.05\) is observed, followed by a steep rise at higher multiplicities corresponding to ultra-central events. This behavior is qualitatively consistent with expectations from hydrodynamic simulations and reflects the increase in temperature with entropy density when the system volume saturates~\cite{CMS:2024kxg}.

Fitting the high-multiplicity region (\(N_{\mathrm{ch}}^{\mathrm{norm}} > 1.14\)) yields~\cite{CMS:2024kxg}

\[
c_s^2 = 0.241 \pm 0.002\ (\mathrm{stat}) \pm 0.016\ (\mathrm{syst}),
\]

with an effective temperature \(T_{\mathrm{eff}} = \langle p_{\mathrm{T}}\rangle^{0}/3 = 219 \pm 8\ (\mathrm{syst})\ \mathrm{MeV}\) (statistical uncertainty is negligible). The relationship \(T_{\mathrm{eff}} \approx \langle p_{\mathrm{T}}\rangle/3\) is motivated by hydrodynamic simulations~\cite{Gardim:2019brr,Gardim:2019nnr} and has been verified in multiple model calculations.

\begin{figure}[!ht]
\begin{center}
\includegraphics[width=0.80\textwidth]{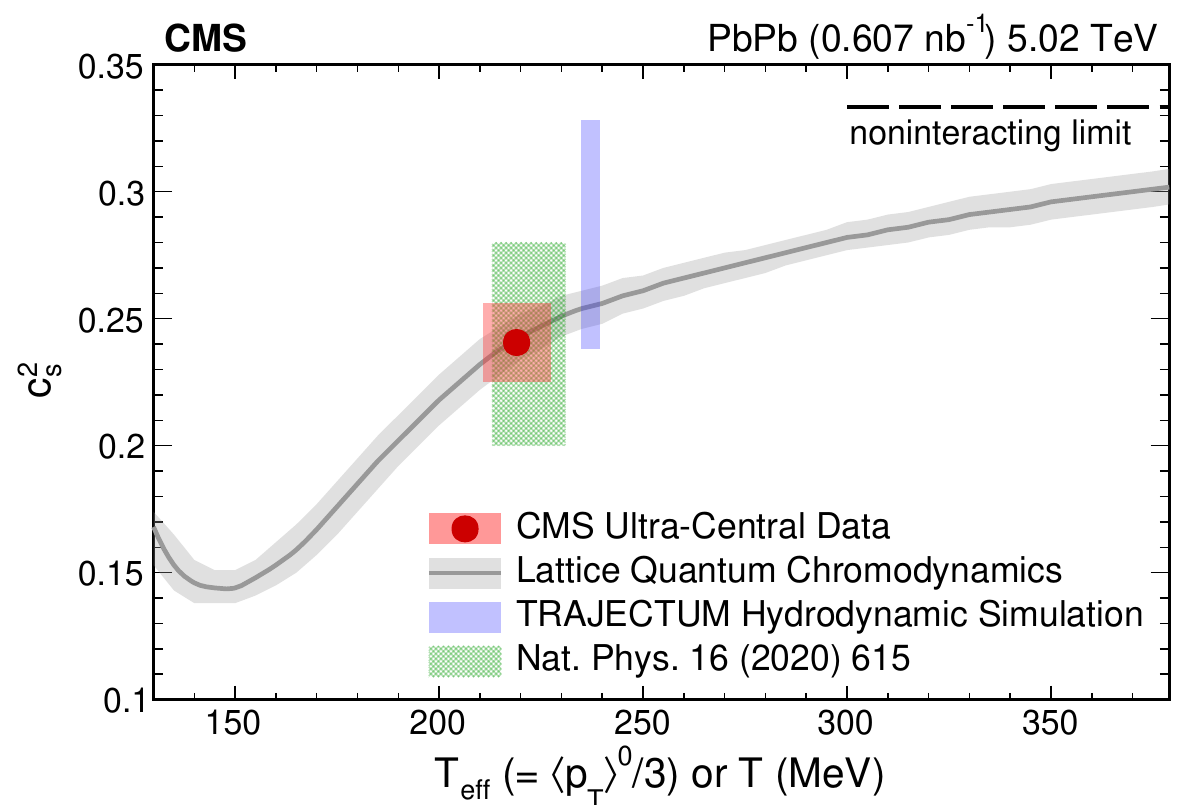}
\caption{The speed of sound squared as a function of the effective temperature. Results for simulations and lattice QCD calculations are also shown~\cite{CMS:2024kxg}.}
\label{fig:2}
\end{center}
\end{figure}

Figure~\ref{fig:2} compares this result with lattice QCD predictions, the TRAJECTUM model, and the previous extraction by Gardim et al.~\cite{CMS:2024kxg}. The CMS measurement shows excellent agreement with lattice QCD, with comparable uncertainties, indicating the formation of a deconfined QCD phase at LHC energies. The TRAJECTUM model yields \(c_s^2 = 0.283 \pm 0.045\) when the same fitting procedure is applied, consistent with data within uncertainties.

\subsection{Studies in pPb collisions}

\begin{figure}[!ht]
\begin{center}
\includegraphics[width=0.80\textwidth]{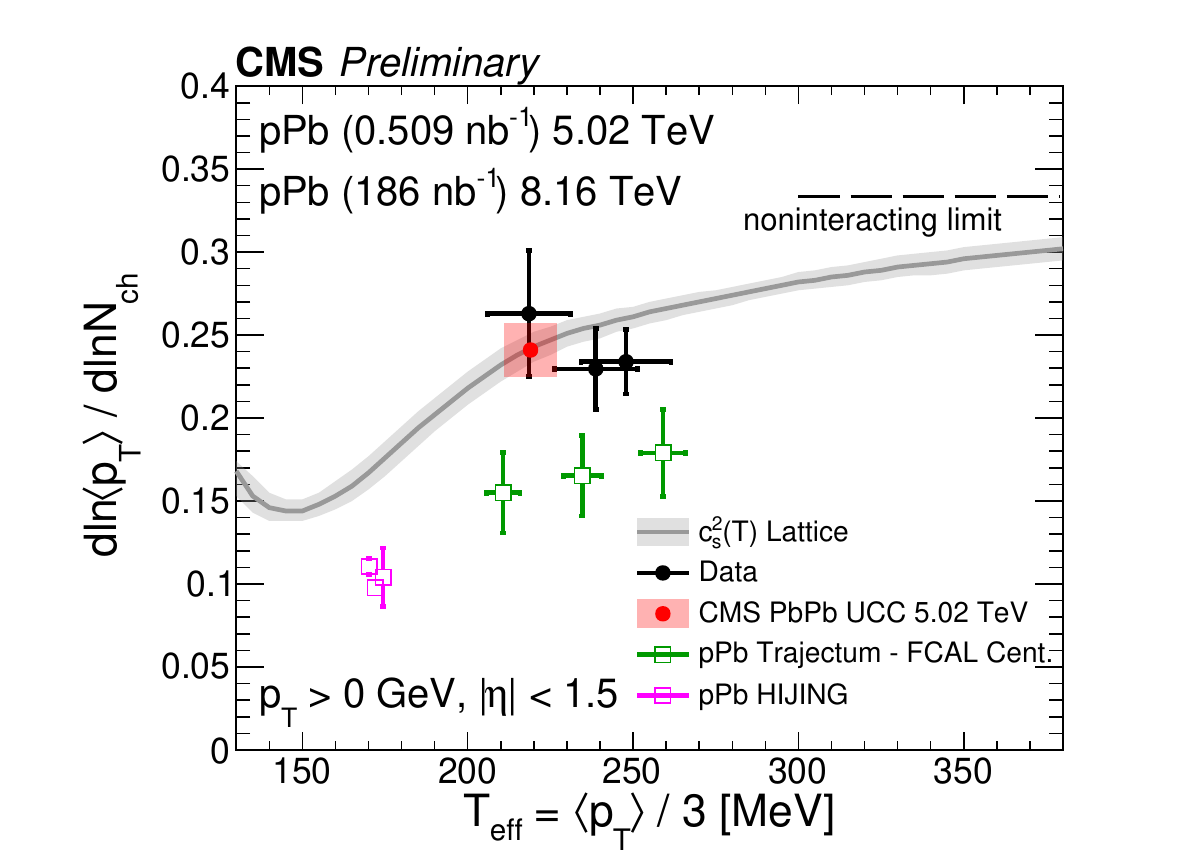}
\includegraphics[width=0.80\textwidth]{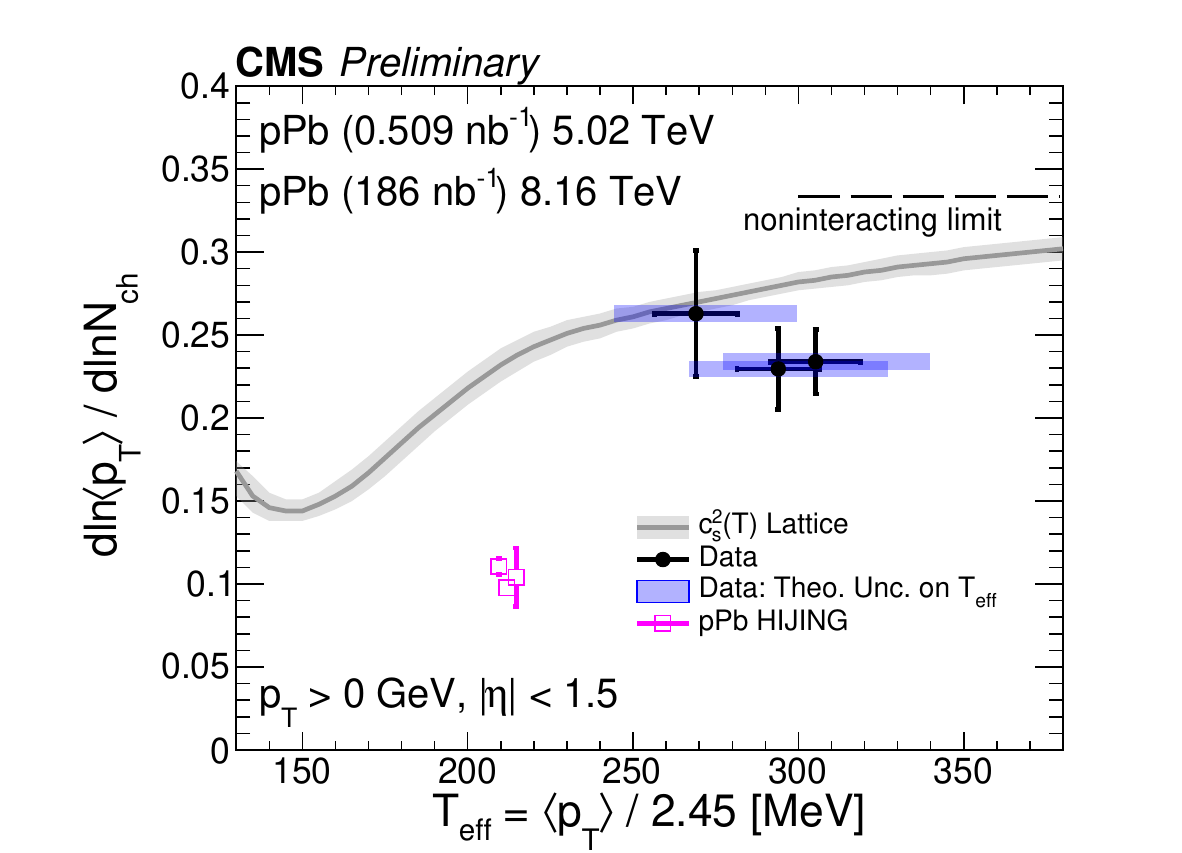}
\caption{The quantity \(\mathrm{d}\ln\langle p_{\mathrm{T}}\rangle/\mathrm{d}\ln N_{\mathrm{ch}}\) is shown as a function of the effective temperature, $T_{\mathrm{eff}}=\langle p_{\mathrm{T}}\rangle/X$, where \(X=3\) (upper) is used for boost-invariant calculations and \(X=2.45\) (lower) for calculations that account for system asymmetries. Comparisons with lattice-QCD calculations and with HIJING and TRAJECTUM simulations are also shown~\cite{CMS:HIN25001}.}
\label{fig:3}
\end{center}
\end{figure}

Figure~\ref{fig:3} presents \(\mathrm{d}\ln\langle p_{\mathrm{T}}\rangle/\mathrm{d}\ln N_{\mathrm{ch}}\) as a function of \(T_{\mathrm{eff}}\) for pPb collisions, compared with lattice QCD, PbPb results, and model calculations~\cite{CMS:HIN25001}. Two scenarios for \(T_{\mathrm{eff}}\) estimation are considered: a boost-invariant picture with \(T_{\mathrm{eff}} \approx \langle p_{\mathrm{T}}\rangle/3\), and a three-dimensional evolution accounting for system asymmetry with \(T_{\mathrm{eff}} \approx \langle p_{\mathrm{T}}\rangle/2.45\).

In the boost-invariant scenario, pPb results agree well with lattice QCD and PbPb data, extending the temperature coverage to a broader range. The agreement suggests that thermodynamic descriptions may be applicable to pPb collisions, particularly in high-multiplicity events where collective effects have been observed. Results from the TRAJECTUM model lie systematically below the data by 2--3 standard deviations.

With the three-dimensional evolution scenario, agreement with lattice QCD worsens, with data lying 1--2 standard deviations below predictions. This highlights the sensitivity of the interpretation to assumptions about the system's dynamical evolution.

\begin{figure}[!ht]
\begin{center}
\includegraphics[width=0.80\textwidth]{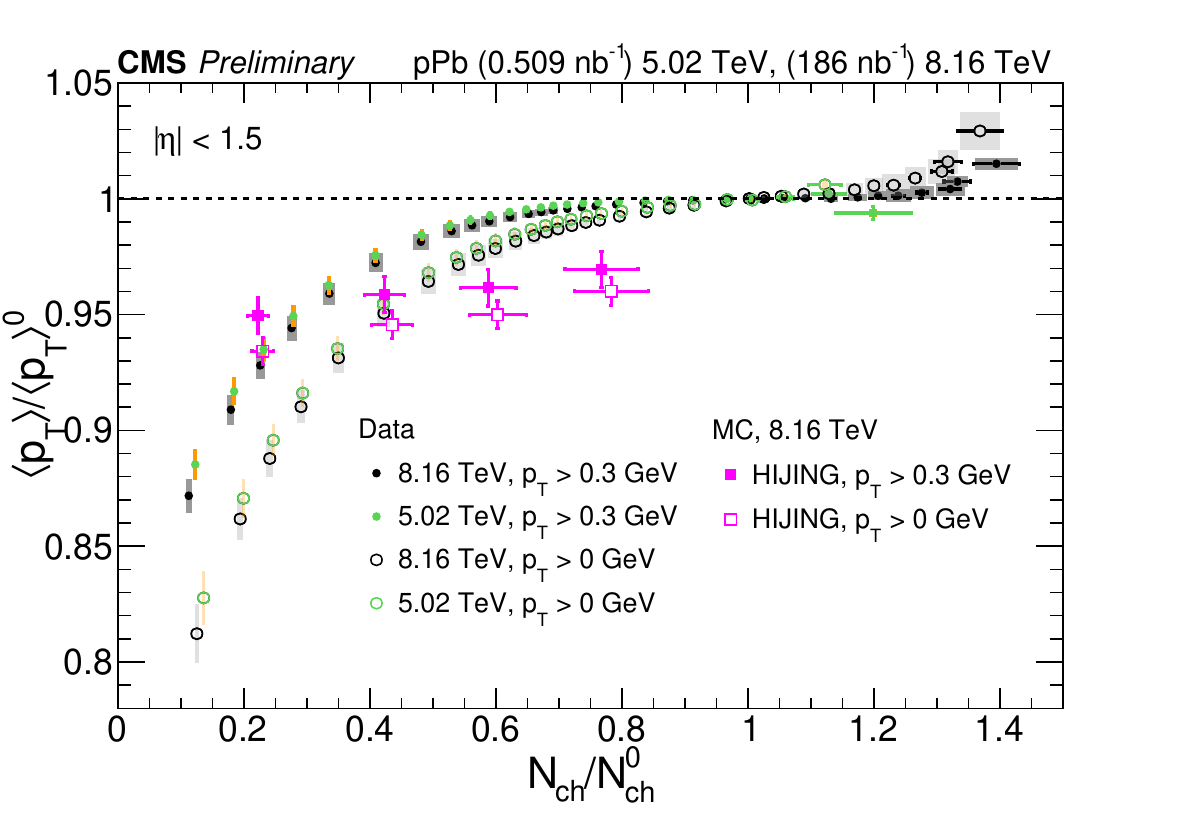}
\caption{The mean transverse momentum as a function of charged-particle multiplicity, with both quantities normalized to their reference values in the 0--0.01\% centrality interval. The results are compared with HIJING simulations~\cite{CMS:HIN25001}.}
\label{fig:4}
\end{center}
\end{figure}

Figure~\ref{fig:4} shows the normalized \(\langle p_{\mathrm{T}}\rangle\) as a function of normalized \(N_{\mathrm{ch}}\) up to the highest multiplicities in pPb collisions~\cite{CMS:HIN25001}. For \(\sqrt{s_{\mathrm{NN}}} = 8.16\ \mathrm{TeV}\), an indication of a rising trend at the highest multiplicities (approximately the top \(10^{-5}\) percentile in centrality) is observed, reminiscent of the behavior in ultra-central PbPb collisions but much less pronounced. The HIJING model fails to describe the data both quantitatively and qualitatively, suggesting that key physical mechanisms in high-multiplicity pPb events are not captured by this non-thermal Monte Carlo.

\section{Summary and outlook}

The CMS experiment has performed precise measurements of the multiplicity dependence of the mean transverse momentum in ultra-central PbPb collisions at \(\sqrt{s_{\mathrm{NN}}} = 5.02\ \mathrm{TeV}\), extracting a squared speed of sound \(c_s^2 = 0.241 \pm 0.002\ (\mathrm{stat}) \pm 0.016\ (\mathrm{syst})\) at \(T_{\mathrm{eff}} = 219 \pm 8\ (\mathrm{syst})\ \mathrm{MeV}\). The good agreement with lattice QCD calculations suggest the formation of a deconfined QCD phase at LHC energies and demonstrates the power of this new hydrodynamic probe.

First measurements in pPb collisions extend these studies to smaller systems, showing consistency with lattice QCD in certain interpretations while highlighting the need for refined theoretical modeling of system-size effects and the relationship between \(\langle p_{\mathrm{T}}\rangle\) and effective temperature.

Future directions include extending these measurements to other collision systems (e.g., OO, NeNe) to probe geometry and system-size effects, exploring the energy dependence of the slope in PbPb collisions at 2.76 TeV, and developing more sophisticated analyses incorporating fluctuation information (e.g., 2D joint probability distributions of \(\langle p_{\mathrm{T}}\rangle\) and \(N_{\mathrm{ch}}\) fluctuations~\cite{Mu:2025kgl}). These efforts will further constrain the EoS of hot QCD matter and deepen our understanding of the strong interaction under extreme conditions. 
 
\section*{Acknowledgements}

This material is based upon work supported by the São Paulo research foundation (FAPESP) grant 2018/25225-9 and CNPq grant 309962/\\2023-4. Any opinions, findings, and conclusions or recommendations expressed in this material are those of the author(s) and do not necessarily reflect the views of FAPESP and CNPq.


\begin{thebibliography}{}

\bibitem{Shuryak:1978ij}
E. V. Shuryak, Sov. Phys. JETP \textbf{47} (1978), 212.

\bibitem{STAR:2004gwh}
STAR Collaboration, Nucl. Phys. A \textbf{757} (2005), 102.

\bibitem{PHENIX:2004wyw}
PHENIX Collaboration, Nucl. Phys. A \textbf{757} (2005), 184.

\bibitem{BRAHMS:2004vtz}
BRAHMS Collaboration, Nucl. Phys. A \textbf{757} (2005), 1.

\bibitem{PHOBOS:2004eai}
PHOBOS Collaboration, Nucl. Phys. A \textbf{757} (2005), 28.

\bibitem{LandauLifshitz}
L. D. Landau and E. M. Lifshitz, \textit{Fluid Mechanics}, Pergamon Press, New York (1959).

\bibitem{CMS:2024kxg}
CMS Collaboration, Rept. Prog. Phys. \textbf{87} (2024), 077801.

\bibitem{Gardim:2019nnr}
F. G. Gardim, G. Giacalone, and J.-Y. Ollitrault, Phys. Lett. B \textbf{809} (2020), 135749.

\bibitem{Gardim:2019brr}
F. G. Gardim, G. Giacalone, M. Luzum, and J.-Y. Ollitrault, Nature Phys. \textbf{16} (2020), 615.

\bibitem{CMS:HIN25001}
CMS Collaboration, CMS Physics Analysis Summary HIN-25-001 (2025).

\bibitem{CMS:2008xjf}
CMS Collaboration, JINST \textbf{3} (2008), S08004.

\bibitem{Lokhtin:2008xi}
I. P. Lokhtin et al., Comput. Phys. Commun. \textbf{180} (2009), 779.

\bibitem{Agostinelli:2002hh}
GEANT4 Collaboration, Nucl. Instrum. Meth. A \textbf{506} (2003), 250.

\bibitem{Hagedorn:1983wk}
R. Hagedorn, Riv. Nuovo Cim. \textbf{6N10} (1983), 1.

\bibitem{Das:2017ned}
S. J. Das, G. Giacalone, P.-A. Monard, and J.-Y. Ollitrault, Phys. Rev. C \textbf{97} (2018), 014905.

\bibitem{Gavassino:2025myk}
L. Gavassino, H. Hirvonen, J.-F. Paquet, M. Singh, and G. Soares Rocha, Phys. Rev. C \textbf{112} (2025), 054903.

\bibitem{Mu:2025kgl}
Y.-S. Mu, J.-A. Sun, L. Yan, and X.-G. Huang, Phys. Rev. Lett. \textbf{135} (2025), 162301.


\end{thebibliography}
\end{document}